\documentclass[lettersize,journal]{IEEEtran}
\usepackage{amsmath,amsfonts}
\usepackage{algorithm}
\usepackage{array}
\usepackage[caption=false,font=normalsize,labelfont=sf,textfont=sf]{subfig}
\usepackage{textcomp}
\usepackage{stfloats}
\usepackage{url}
\usepackage{verbatim}
\usepackage{graphicx}
\usepackage{cite}
\usepackage{algorithmicx}
\usepackage{algpseudocode}
\usepackage{textgreek}
\usepackage{soul}
\soulregister\ref7
\usepackage{color}

\begin{document}

    \title{Fundamental Techniques for Optimal Control of Reconfigurable Battery Systems: System Modeling and Feasible Search Space Construction}
    \author{Changyou~Geng,
        ~Dezhi~Ren,~\IEEEmembership{Student Member},~\IEEEmembership{IEEE},
        ~Enkai~Mao,~\IEEEmembership{Student Member},~\IEEEmembership{IEEE},
        ~Xinyi~Zheng,~\IEEEmembership{Student Member},~\IEEEmembership{IEEE},
        ~Mario~Va\v{s}ak,~\IEEEmembership{Senior Member},~\IEEEmembership{IEEE},
        ~Changfu~Zou,~\IEEEmembership{Member},~\IEEEmembership{IEEE},
        ~Weiji~Han,~\IEEEmembership{Member},~\IEEEmembership{IEEE}
    }

    \markboth{Journal of XXX,~Vol.~XX, No.~XX, January~2025}%
    {Shell \MakeLowercase{\textit{et al.}}: A Sample Article Using IEEEtran.cls for IEEE Journals}

    \maketitle

    \begin{abstract}

        Reconfigurable battery systems (RBSs) are emerging as a promising solution to improving fault tolerance, charge and thermal balance, energy delivery, etc. To optimize these performance metrics of RBSs, high-dimensional nonlinear integer programming problems need to be formulated and solved. To accomplish this, it is necessary to address several critical challenges stemming from nonlinear battery characteristics, discrete switch states, dynamic system configurations, as well as the curse of dimensionality inherent in large-scale RBSs. Thus, we propose a unified modeling framework to accommodate various possible configurations of an RBS and even to cover different RBS designs and their hybrid combinations, enabling the problem formulation for the RBS optimal control and facilitating the RBS topology design.
        Further, to solve the formulated RBS optimal control problems, the search space is narrowed to encompass only the feasible solutions, thereby ensuring safe battery connections while substantially curtailing search efforts. These proposed techniques, focusing on unifying the system modeling and narrowing the search space, lay a solid foundation for effectively formulating and efficiently solving RBS optimal control problems. The accuracy and effectiveness of the proposed techniques are demonstrated by both simulation and experimental tests.
    \end{abstract}


    \section{Introduction}
    \label{sec:introduction}
    \IEEEPARstart{I}{}n response to the urgent imperative of mitigating greenhouse gas emissions, there has been a notable surge in the adoption of renewable energy sources, such as solar panels and wind turbines \cite{yang2022Assessment}. Batteries, serving as a primary energy storage medium, play a pivotal role in both integrating renewable energy into power grids and facilitating its application to transportation electrification \cite{xinjing2023Economic}.
    Given the intrinsic electrochemical nature of batteries, multiple batteries need to be connected either in series or in parallel to meet diverse voltage, current, or power demands across various applications. For multi-battery systems, inconsistent battery parameters and unbalanced operating states, such as state of charge (SoC), temperature, and state of health (SoH), result in a range of issues spanning from diminished energy delivery \cite{lu2013review}, accelerated degradation \cite{barreReviewLithiumionBattery2013a} to potential safety hazards \cite{xuning2018Thermal}.
    Addressing these issues arising from unbalanced battery operation involves manipulating the operating currents of relevant batteries. However, in many practical battery systems, implementing current control of each individual cell is infeasible due to the commonly fixed battery connection configuration.

\begin{figure}[ht]
        \vspace{-15pt}
        \centering
        \includegraphics[width=\columnwidth]{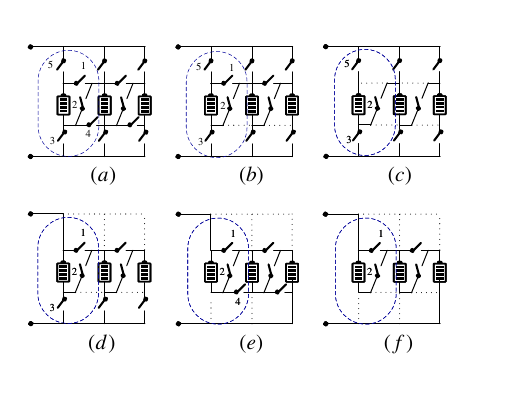}
        \vspace{-30pt}
        \caption{Some structure designs for RBSs proposed in recent literature.}
        \label{fig:RBS_designs}
    \end{figure}

    Aimed at mitigating the limitations imposed by fixed battery connections on system performance control, reconfigurable battery systems (RBSs), employing switches to dynamically reconstruct battery connections, are emerging as a promising solution \cite{song2016Reconfigurable}.
    Various RBS designs, featuring diverse switch arrays and demonstrating different levels of reconfigurability, have been proposed \cite{taesic2011Seriesconnected,  younghyun2011Balanced,  sebastian2016Distributed,horacio2008Reconfigurable, mahmoud2008Battery, song2012Dynamic}. For instance, Fig. \ref{fig:RBS_designs} illustrates some RBS structure designs with two to five switches associated with each battery cell. Through switch operations, the battery connection can be reconfigured, enabling the recombination of battery terminal voltages and redistribution of currents among batteries. Then, by designing and implementing appropriate control algorithms, one can potentially improve the RBS's operation performance in terms of enhanced fault tolerance \cite{mahmoud2008Batterya}, extended energy delivery \cite{zhiliang2015Distributed}, balanced battery states \cite{weiji2019Fastest} and thermal distribution \cite{faisal2017Load}, and greater power efficiency \cite{2012Power}. Moreover, RBSs can accommodate batteries of heterogeneous age, type and even chemistry \cite{anirudh2015Software}, and facilitate the diagnosis, repair, and replacement of local cells as well.

    In order to fully leverage these potential benefits of RBSs, we need to formulate one specific optimal control problem, typically consisting of an objective function, a system model, and some necessary constraints, as will be detailed in Section \ref{sec:RBS_optimal_control_formulation}. The system model describing the system state evolution is formulated based on the electrical interconnection of all batteries, i.e., the \textit{system configuration}. The system configuration of an RBS can be specified by a switch state vector (SSV) composed of the on-off states of all switches involved in the system. The evolution of system states and the value of objective function are directly influenced by the specified system configuration. Motivated by this, the RBS's configuration can be selected as one control variable. Through appropriate reconfiguration, various system performance metrics defined in the objective function are expected to be optimized.    

    To formulate and solve an optimal control problem for the RBS, a critical challenge is how to construct a system model to simulate the dynamically varying system configuration, which has been rarely investigated in the literature.
    As the control variable, the RBS's configuration could vary within a wide range. The optimal system configuration is typically approached through an iterative evaluation-and-update process. During each iteration, a batch of configurations (or configuration series if multiple decision steps are involved) are first evaluated and then updated according to certain principles.
    To evaluate the performance of a specific configuration, we need to construct its corresponding system model based on all battery cells' mathematical models as well as equations characterizing their electric connection according to Kirchhoff's current law (KCL) or voltage law (KVL) \cite{weiji2020NextGeneration}. When evaluating another system configuration of the RBS, in which some battery cells quit or join the operation and certain local series or parallel connections might change as well, it becomes difficult to directly revise the original system model since some entry values and even the dimensions of system matrices might change. In this case, it is commonly necessary to construct a new system model. However, it is challenging to generate new models automatically by programming codes due to involving re-indexing the batteries put into operation, rewriting their KCL and KVL equations, and reformulating the system matrices accordingly. Consequently,  manually constructing different models for evaluating various system configurations becomes very inefficient. 
   
   To address the issues caused by dynamic reconfiguration, we propose the concept of \textit{complete system configuration} (CSC) to unify the modeling of various RBS configurations in Section \ref{sec:unified-modeling-framework}. The formulated system model based on the CSC, explicitly expressed in terms of the system configuration or SSVs, is capable of covering all possible configurations of an RBS without reconstruction, significantly reducing the computational efforts of evaluating various system configurations. Thanks to this modeling approach, a single model constructed for design (a) in Fig. \ref{fig:RBS_designs} can represent all other designs shown in the figure, as well as hybrid combinations of these designs.

    When solving an RBS optimal control problem, another critical challenge is how to efficiently search the optimal solution. During the process of iterative evaluation and update, the number of possible system configurations experiences exponential growth with the number of switches involved in the RBS as well as the number of decision steps within the decision horizon, as will be demonstrated in Section \ref{sec:construction-of-narrowed-search-space}.
    Thus, for large-scale RBSs, it becomes very time-consuming to search for the optimal solution among a huge number of possible system configurations.

    To alleviate the influence of the ``curse of dimensionality'', we will propose an approach to actively construct the feasible solution space in Section \ref{sec:construction-of-narrowed-search-space}. By filtering all SSVs incurring open circuits, short circuits, undesired system voltages, or duplicate configurations, the size of the constructed feasible solution space is much smaller than that of the complete solution space exponentially expanding with the number of switches. In addition to ensuring safe operation, the proposed approach for narrowing the search space can dramatically enhance the search efficiency when solving optimal control problems for RBSs.

    The remainder of this work is organized as follows. First, the general optimal control problem for RBSs is formulated in Section \ref{sec:RBS_optimal_control_formulation}.
    In Section \ref{sec:unified-modeling-framework}, the unified modeling framework for RBSs is introduced along with simulation and experimental tests to demonstrate the modeling accuracy.
    Then, the approach to narrow the search space is investigated in Section \ref{sec:construction-of-narrowed-search-space}, and its performance is compared with conventional methods.
    Finally, conclusions are presented in Section \ref{sec:conclusion}.

    \section{Formulation of a general optimal control problem for RBSs}\label{sec:RBS_optimal_control_formulation}
    A general optimal control problem is formulated as follows.
    \begin{equation*}
        \begin{aligned}
            &\min ~ J(X_0, U, D) \\
            \text{s.t.} ~ &X_{k} = \Phi(X_{k-1}, u_k, D_k), \quad k=1,2,\ldots,K,\\
            &g_l(X_0, U, D)\geq 0, \quad l=1,2,\ldots,L,\\
            &h_r(X_0, U, D)=0, \quad r=1,2,\ldots,R,\\
            &U\in \mathcal{U}^K.
        \end{aligned}
    \end{equation*}

    In the objective function $J$, $X_k$ denotes the system state vector at $t=k\Delta t$, $U = [u_1,u_2,\ldots,u_K]$ the input trajectory, and $D = [D_1,D_2,\ldots,D_K]$ the disturbances. Notation $\Delta t$ is the duration of each sampling interval and $K$ is the number of sampling intervals considered during the decision horizon. Then, $X_0$ is the initial system state vector.
    Among the constraints, the first one describes the system model, followed by some inequality and equality constraints, and $\mathcal{U}^K$ denotes the constraint set for the input trajectory.

    Different from a traditional battery system, RBS enables the possibility to select the system configuration, described by switch state vectors (SSVs), as one control variable or system input for optimal control. Specifically, for the $k$-th sampling interval $((k-1)\Delta t, k\Delta t]$, the corresponding input $u_k$ is expressed as follows based on the on-off state of each switch, denoted by $s_{n,k} \in \mathcal{B} = \{0,1\}$, $n=1,2,\ldots, N_S$, where $N_S$ is the total number of switches in the RBS.
    \begin{equation}
       \label{eq:SSV}
        \begin{aligned}
            u_k= [ s_{1,k},s_{2,k},\ldots, s_{N_S,k}]\in \mathcal{B}^{N_S}.
        \end{aligned}
    \end{equation}
    Moreover, the system's load profile, commonly specified in terms of system power or current, is regarded as an external disturbance $D$ in the above problem.
    
\begin{figure}[ht]
        \centering
        \includegraphics[width=\columnwidth]{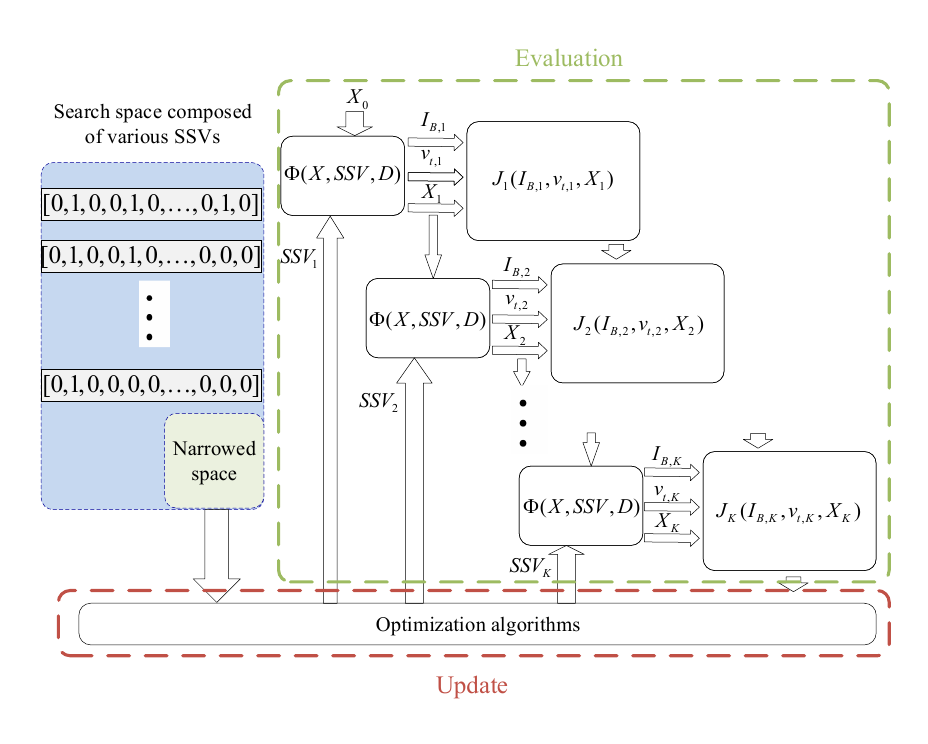}
            \vspace{-30pt}
        \caption{Structural diagram of the proposed framework.}
        \label{fig:structure}
    \end{figure}

    The general process of iterative evaluation and update of SSVs for searching the optimal solution is illustrated in Fig. \ref{fig:structure}, in which the right part describes the performance evaluation of SSVs, the left part presents the search space of SSVs, and the optimization algorithms at the bottom generates the updated SSVs for each iteration. Among these three parts, the first two parts lay the foundation for adopting various optimization algorithms. Thus, this work is mainly focused on boosting the efficiency of evaluating and searching SSVs.

    For efficiently evaluating the RBS performance metric described by the objective function $J$, in Section \ref{sec:unified-modeling-framework}, a unified system model with the SSV as input is developed to accommodate diverse system configurations, corresponding to the unified state transition function $ \Phi(X,SSV,D) $ shown in Fig. \ref{fig:structure}. The objective function $J$ depends on the vector of all battery cell currents $I_B$, the system's terminal voltage $v_t$, and the system's state variable vector $X$.   
    To further improve the searching efficiency, instead of directly searching within the complete search space $\mathcal{B}^{N_S}$, we will identify only those feasible SSVs to form a dramatically narrowed search space, i.e., the feasible search space, in Section \ref{sec:construction-of-narrowed-search-space}.

    \section{Unified modeling framework for RBSs}\label{sec:unified-modeling-framework}
    Different structures can be designed for RBSs by changing the number of switches around each battery cell and the cell-switch connection patterns. For instance, as shown in Fig. \ref{fig:RBS_designs}, RBS design (a) can be used to realize other designs (b)–(f) by selectively removing specific switch or wire branches (i.e., disconnecting the corresponding switches). Thus, we will take RBS design (a) as an example to introduce the unified modeling framework for RBSs.

    \subsection{Complete system configuration for unifying the RBS modeling}\label{subsec:complete_system_configuration}
\begin{figure}[ht]
            \centering
            \includegraphics[width=\columnwidth]{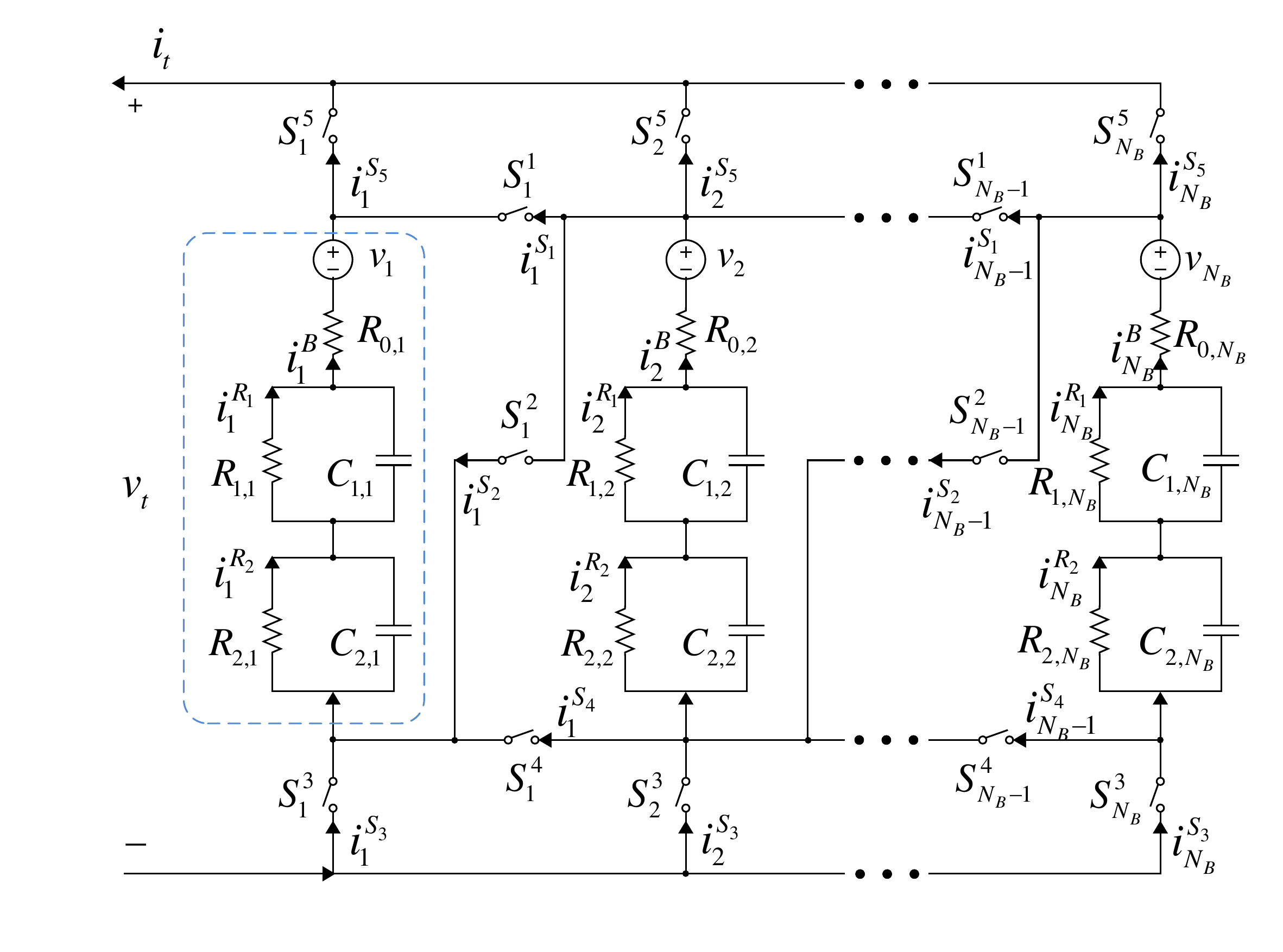}
            \vspace{-25pt}
            \caption{Diagram for RBS design (a) in Fig. \ref{fig:RBS_designs}.}
            \label{fig:circuit}
    \end{figure}

    To model an RBS, not only all components, such as batteries, switches, wires, and contactors, but also the \textit{system configuration}, i.e., how all components are interconnected with each other, needs to be described. The diagram for RBS design (a) is presented in Fig. \ref{fig:circuit}, in which the $n$-th battery cell, $n=1,2,\ldots,N_B$, is represented by a second-order equivalent circuit model (ECM) \cite{stevenc.1993Simple}, and $N_B$ is the total number of battery cells. In this battery cell model, $v_n$ denotes the open circuit voltage (OCV), $R_{0,n}$ denotes the internal resistance, and two parallel resistor-capacitor (RC) pairs are used for describing the lithium diffusion process. Other types of battery models can be deployed as well if necessary. The five switches associated with the $n$-th battery cell are labeled as $S^m_n$, $m=1,2,\ldots,5$, and each switch's state is denoted by $s^m_n\in\{0,1\}$. If a switch's state is on, i.e., conducted, $s^m_n=1$ and its resistance is represented by its static drain-to-source on resistance $R_{\text{ds(on)},n}^{m}$. Otherwise, the off switch state $s^m_n=0$ indicates that the corresponding branch is disconnected.

    Clearly, an RBS's configuration depends on the on or off switch states. 
    If those branches including any off-state switches are removed from the diagram in Fig. \ref{fig:circuit}, some local meshes are changed accordingly. Consequently, we need to reconstruct the system model since some battery cells need to be re-indexed, and their corresponding KCL or KVL equations need to be rewritten as well. It is challenging to do so automatically by programming codes. Thus, it becomes very inefficient to model the RBS under dynamic reconfiguration.

To boost the RBS modeling efficiency by avoiding frequent model reconstruction, it is better to keep all meshes unchanged despite the dynamic reconfiguration. Thus, instead of directly removing those off-state switch branches, all branches are reserved so that we can maintain a \textit{complete system configuration} (CSC) involving all switches, batteries, and wire branches. 
The CSC features stable meshes regardless of any reconfiguration, significantly facilitating the formulation of KCL and KVL equations. As a result, it becomes possible to develop a unified system model to cover all configurations of an RBS.

When modeling the CSC of an RBS, each switch is characterized by a switch resistance relying on the switch state.
Then, the resistance of switch $S^m_n$ is denoted by $R^{S_m}_{n}$ and expressed through the binary switch state $s^m_n$ as follows. 
\begin{align}
	R^{S_m}_{n} = s^m_n R_{\text{\text{ds(on)}},n}^{m} +(1-s^m_n) R_{\text{ds(off)},n}^{m}+ R_{\text{wire},n}^{m}.\label{eq:rds}
\end{align}
In this expression, $R_{\text{ds(off)},n}$ denotes the switch's drain-to-source off resistance, which is typically at M$\Omega$ level for practical MOSFETs \cite{Vishay2016_N}.
Moreover, to simplify the notation, the resistance of the connection wire along the switch branch, denoted by $R_{\text{wire},n}^{m}$, is incorporated into $R^{S_m}_{n}$. According to (\ref{eq:rds}),  the switch branch's resistance $R^{S_m}_{n}=R_{\text{\text{ds(on)}},n}^{m}+ R_{\text{wire},n}^{m}$ if the switch is on, or $R^{S_m}_{n}=R_{\text{\text{ds(off)}},n}^{m}+ R_{\text{wire},n}^{m}$ if the switch is off. Then, during dynamic reconfiguration of an RBS, the switch resistance needs to be updated if its state is changed but the mathematical expression of the CSC model can always remain unchanged. 

\subsection{Formulation of an RBS model based on its CSC}\label{subsec:rbs-model}

To formulate the state-space representation for the RBS model based on its CSC, the states of battery cells shown in Fig. \ref{fig:circuit}, including each cell's resistor currents $i_n^{R_1}$ and $i_n^{R_2}$ in the RC pairs and the OCV $v_n$, are chosen to compose the RBS state vector as follows. 
\begin{equation}
        \label{eq:state_total}
        X = [V^T, I_{R_1}^T, I_{R_2}^T]^T.
\end{equation}
In the above state vector, $V=[v_1,v_2,\dots,v_{N_B}]^T$ is the vector of battery cell OCVs, and $I_{*} = [i^{*}_1,i^{*}_2,\ldots,i^{*}_{N_B}]^T$ denotes the current vector of all components specified by the wildcard $*$. For instance, $I_B$, $I_{S_m}$, and $I_{R_1}$ represent the current vectors of all battery cells, all the $m$-th switches around battery cells, and all $R_1$'s in the battery cell ECMs, respectively.
In addition, the system's total current $i_t$ is regarded as the system disturbance. The system's output can be selected according to the objective function $J$.   

For the $n$-th battery cell, the current $i_n^B={\mathrm{d}Q_n}/{\mathrm{d}t}$, $n=1, 2,\ldots, N_B$, where $Q_n$ denotes the cell's amount of charge. The $n$-th cell's SoC is represented with $z_n={Q_n}/{Q_{C,n}}$, where $Q_{C,n}$ denotes its charge capacity. According to the $n$-th cell's OCV-SoC curve, the $n$-th cell's OCV denoted by $v_n$ can be regarded as a function of cell SoC $z_n$, and hence the slope of $v_n$ at any specific $z_n$ is defined by $k_n^V(z_n)=\frac {\mathrm{d}v_n}{\mathrm{d}z_n}$. Then, the state-space equation for each cell's OCV is described by 
\begin{equation}\label{eq: mV}
\dot v_n=\frac {\mathrm{d}v_n}{\mathrm{d}t}=\frac {\mathrm{d}v_n}{\mathrm{d}z_n}\frac{\mathrm{d}z_n}{\mathrm{d}t}=\frac {k_n^{V}(z_n)}{Q_{C,n}}\frac{\mathrm{d}Q_n}{\mathrm{d}t}=\frac {k_n^{V}(z_n)}{Q_{C,n}}i_n^B,
\end{equation}
and the corresponding matrix form for the OCV vector is 
\begin{equation}\label{eq{state_OCV}}
    \dot V=B_VI_B.
\end{equation}

To derive the state-space equations describing the evolution of $I_{R_1}$ and $I_{R_2}$ in the state vector $X$, KCL and KVL equations need to be formulated based on the CSC in Fig. \ref{fig:circuit}, where  
$i^{S_m}_{n}$ denotes the current of the $m$-th switch around the $n$-th cell, $m=1,2,\ldots,5$, $n=1,2,\ldots,N_B$. 
The KCL equations at the positive and negative electrodes of the n-th cell are formulated as follows.
\begin{align}
        \label{eq:o_KCL_p}
        &\begin{cases}
        -i^B_{n} -i_n^{S_1} +i^{S_5}_{n}=0,  & \text{if }  n=1,\\
         -i^B_{n}-i_n^{S_1}+i^{S_5}_{n}  +i_{n-1}^{S_1} +i^{S_2}_{n-1} =0, &  \text{if } 1<n<N_B,\\
         -i^B_{n}+i^{S_5}_{n}  +i_{n-1}^{S_1} +i^{S_2}_{n-1} =0, &  \text{if } n=N_B.
        \end{cases}
\\
        \label{eq:o_KCL_n}
        &\begin{cases}
        i^B_{n}-i^{S_2}_{n}-i^{S_3}_{n} - i^{S_4}_{n} =0,  & \text{if }  n=1,\\
        i^B_{n}-i^{S_2}_{n}-i^{S_3}_{n} - i^{S_4}_{n} + i^{S_4}_{n-1} =0,  & \text{if }  1<n<N_B,\\
        i^B_{n}-i^{S_3}_{n} + i^{S_4}_{n-1} =0, &  \text{if } n=N_B.
        \end{cases}
    \end{align}
At the positive and negative terminals of the RBS, the KCL equations are formulated as follows.
\begin{equation}
        \label{eq:teminal_p}
        \sum_{n=1}^{N_B} i^{S_5}_{n} = i_t,
    \end{equation}
    \begin{equation}
        \label{eq:teminal_n}
        \sum_{n=1}^{N_B} i^{S_3}_{n} = i_t.
    \end{equation}

For the four meshes between the $n$-th and $(n+1)$-th cell, $1\leq n<N_B$, the KVL equations are formulated as follows.  
    \begin{align}
    \label{eq:mesh1}
    &R^{S_5}_{n}i^{S_5}_{n} - R^{S_5}_{n+1}i^{S_5}_{n+1} + R^{S_1}_{n}i_n^{S_1} = 0, \\
    \label{eq:mesh2}
    &R_{0,n}i^B_{n} - v_n - R^{S_1}_{n}i_n^{S_1} + R^{S_2}_{n}i^{S_2}_{n} + R_{1,n}i^{R_1}_{n} + R_{2,n}i^{R_2}_{n} = 0, \\
    \label{eq:mesh3}
    &R_{0,n+1}i^B_{n+1} - v_{n+1} + R_{1,n+1}i^{R_1}_{n+1} + R_{2,n+1}i^{R_2}_{n+1} + R^{S_2}_{n}i^{S_2}_{n} \notag \\
    &- R^{S_4}_{n}i_n^{S_4} = 0, \\
    \label{eq:mesh4}
    &R^{S_3}_{n}i^{S_3}_{n} - R^{S_3}_{n+1}i^{S_3}_{n+1} - R^{S_4}_{n}i_n^{S_4} = 0.
\end{align}

In the above KCL and KVL equations, it is aimed to cancel all intermediate variables, $i_n^B$, $i_n^{S_m}$, $n=1,2,\dots,N_B$, $m=1,2,\dots,5$, so that only state variables in $X$ and the system input $i_t$ are left in the final state-space equations.
At first, intermediate variables $i_{n}^{S_1}$ and $i_{n}^{S_4}$, $1\leq n<N_B$, 
can be expressed as (\ref{eq:KCL_p}) and (\ref{eq:KCL_n}) based on (\ref{eq:o_KCL_p}) and (\ref{eq:o_KCL_n}), respectively. 
    \begin{equation}
        \label{eq:KCL_p}
        i_n^{S_1} = -i^B_{1}+i^{S_5}_{1}+\sum_{j=2}^{n} (-i^B_{j}+i^{S_5}_{j}+i^{S_2}_{j-1}),
    \end{equation}
\begin{equation}
        \label{eq:KCL_n}
        i_n^{S_4}=\sum_{j=1}^{n} (i^B_{j}-i^{S_2}_{j}-i^{S_3}_{j}).
\end{equation}

After substituting all $i^{S_1}_{n}$ and $i^{S_4}_{n}$ in (\ref{eq:mesh1})-(\ref{eq:mesh4}) using (\ref{eq:KCL_p}) and (\ref{eq:KCL_n}), KVL equations (\ref{eq:mesh1})-(\ref{eq:mesh4}) are combined with KCL equations (\ref{eq:teminal_p}) and (\ref{eq:teminal_n}) and then reorganized into the following matrix forms to facilitate the derivation, where $ E_*, F_*, G_*,  $ and $ H_* $ are $N_B \times N_B$ coefficient matrices of the corresponding current vectors. To have square matrices $F_{S_2}$ and $G_{S_2}$, a virtual switch current $i_{N_B}^{S_2}=0$ is appended to form a $N_B \times 1$ vertical vector $I_{S_2}$.
    \begin{align}
    \label{eq:mesh3_matrix}
    E_{S_5} I_{S_5} &= E_B I_B + E_{S_2} I_{S_2} + E_t i_t, \\
    \label{eq:mesh1_matrix}
    F_{S_2} I_{S_2} &= F_B I_B + F_{R_1} I_{R_1} + F_{R_2} I_{R_2} + F_v V + F_{S_5} I_{S_5}, \\
    \label{eq:mesh2_matrix}
    G_{S_2} I_{S_2} &= G_B I_B + G_{R_1} I_{R_1} + G_{R_2} I_{R_2} + G_v V + G_{S_3} I_{S_3}, \\
    \label{eq:mesh4_matrix}
    H_{S_3} I_{S_3} &= H_B I_B + H_{S_2} I_{S_2} + H_t i_t.
\end{align}
By substituting the $I_{S_5}$ in (\ref{eq:mesh1_matrix}) with the $I_{S_5}$ derived from (\ref{eq:mesh3_matrix}), we can obtain (\ref{eq:matrix1_tidy}). Similarly, the $I_{S_3}$ in (\ref{eq:mesh2_matrix}) can be substituted with the one obtained from (\ref{eq:mesh4_matrix}), leading to (\ref{eq:matrix2_tidy}). In equations (\ref{eq:matrix1_tidy}) and (\ref{eq:matrix2_tidy}), $J_*$ and $K_*$ are computed from $ E_*, F_*, G_*,  $ and $ H_* $. 
 \begin{align}
    J_{S_2} I_{S_2} &= J_B I_B+J_{R_1} I_{R_1} + J_{R_2} I_{R_2} +J_v V+M_t i_t,\label{eq:matrix1_tidy}\\
    K_{S_2} I_{S_2}&=K_B I_B+K_{R_1} I_{R_1} + K_{R_2} I_{R_2} +K_v V+K_t i_t.\label{eq:matrix2_tidy}
\end{align}   
Through combining (\ref{eq:matrix1_tidy}) and (\ref{eq:matrix2_tidy}), $I_{S_2}$ can be eliminated. Then, the cell current vector $I_B$ can be expressed as follows based on the state vector defined in (\ref{eq:state_total}). 
\begin{equation}
        \label{eq:output_I_B}
        I_B=C_{I_B}X+D_{I_B}i_t.
    \end{equation}
 
The derived equation (\ref{eq:output_I_B}) plays an important role in bridging the gap between the battery system-level current $i_t$ and all cell-level currents in $I_B$.
According to the battery cell ECMs in Fig. \ref{fig:circuit}, we have 
    \begin{equation}
        \label{eq:matrix^i_is}
        I_B = I_{R_1}+T_1\dot{I}_{R_1} = I_{R_2}+T_2\dot{I}_{R_2},
     \end{equation}
where $\dot{I_{*}} = {\text {d}I_{*}}/{\text {d}t}$, and $T_1$ and $T_2$ represent $N_B\times N_B$ diagonal matrices with the RC pairs' time constants $\tau_{1,n}=R_{1,n}C_{1,n}$ and $\tau_{1,n}=R_{2,n}C_{2,n}$ as the $n$-th diagonal entries, respectively. Then, combining (\ref{eq:output_I_B}) and (\ref{eq:matrix^i_is}) leads to the following state-space equations characterizing the evolution of $I_{R_1}$ and $I_{R_2}$, respectively.
\begin{align}
 \dot{I}_{R_1} &=A_{I_{R_1}}X + B_{I_{R_1}}i_t,\label{eq:state_I_R1}\\  
 \dot{I}_{R_2} &=A_{I_{R_2}}X + B_{I_{R_2}}i_t.\label{eq:state_I_R2}
\end{align}
        
Finally, based on the defined state vector $X$ in (\ref{eq:state_total}), we can reorganize all state evolution equations in (\ref{eq{state_OCV}}), (\ref{eq:state_I_R1}), and (\ref{eq:state_I_R2}) to formulate the state-space model for the battery system as follows.
\begin{equation}\label{eq:system_plant}
        \dot{X} = A X + B i_t.
\end{equation}
In order to pursue higher computational efficiency and match the discrete-time system model in the formulated optimal control problem at the beginning of Section \ref{sec:unified-modeling-framework}, the state-space model in (\ref{eq:system_plant}) can be further discretized according to, e.g., Zero-Order Holder method or Euler method.  
Considering that some ECM parameter values and the OCV-SoC slope $k_n^V$ of battery cells vary with cell SoCs, they are updated based on the new SoCs and the pre-identified look-up tables at the beginning of every sampling interval. Then, the coefficient matrices $ A $, $ B $, $ C $, and $ D $ in the system model dependent on these cell parameter values are updated accordingly. Thus, battery parameter values and coefficient matrices could change within a decision step consisting of multiple sampling intervals.

Based on the system model formulated in (\ref{eq:system_plant}), we can derive the expression for any system output of interest, such as the cell current vector $I_B$, cell terminal voltage vector denoted by $V_B$, and the system's terminal voltage denoted by $v_t$ as shown in Fig. \ref{fig:circuit}. $I_B$ is already expressed in (\ref{eq:output_I_B}) in terms of the system's state vector $X$ and input $i_t$. 
Denote the terminal voltage of the $n$-th battery cell by $v^B_n$, $n=1,2,\ldots,N_B$. Then, the cell terminal voltage vector is $V_B=[v^B_1,v^B_2,\ldots,v^B_{N_B}]^T$. According to the battery cell ECM in Fig. \ref{fig:circuit}, 
\begin{align}
v^B_n &= v_n-R_{0,n}i_{n}^B-R_{1,n}i^{R_1}_{n}-R_{2,n}i^{R_2}_{n}. \label{eq:v^B_n} 
\end{align}
Then, $V_B$ can be expressed as follows based on (\ref{eq:state_total}), (\ref{eq:output_I_B}), and (\ref{eq:v^B_n}).
\begin{equation}\label{eq:output_V_B}
    V_B=C_{V_B}X+D_{V_B}i_t.
\end{equation}
In Fig. \ref{fig:circuit}, the system's terminal voltage $v_t$ can be calculated by
\begin{align}
v_t&=v^B_n-R^{S_3}_{n}i^{S_3}_{n}-R^{S_5}_{n}i^{S_5}_{n}.\label{eq:v_t}
\end{align}
Specify one battery cell $n$ in (\ref{eq:v_t}), and then based on (\ref{eq:mesh3_matrix}), (\ref{eq:mesh4_matrix}), (\ref{eq:matrix1_tidy}), (\ref{eq:output_I_B}), and (\ref{eq:output_V_B}), the battery system's terminal voltage can be expressed as follows
\begin{equation}
        \label{eq:output_v_t}
        v_t = C_{v_t} X + D_{v_t} i_t.
\end{equation}
Due to the variability of the terminal voltage in RBSs, the input current is influenced not only by the load power but also by the system configuration.
Here, we present a simple and accurate method for determining the input current, with its accuracy demonstrated in Section  \ref{subsec:model-verification}.
Based on (\ref{eq:output_v_t}), given a power request $p_t$ during each sampling interval in practical applications of battery systems, the system's input current $i_t$ can be calculated by solving the quadratic equation
\begin{equation}
        \label{eq:power2current}
        p_t=v_t i_t=C_{v_t} X i_t + D_{v_t} i^2_t.
\end{equation}

During the above formulation of a unified RBS state-space model, it is worth noting that the specified SSV for each sampling interval determines the switch resistances in the KVL equations (\ref{eq:mesh1})-(\ref{eq:mesh4}) through the proposed switch resistance variable in (\ref{eq:rds}), and, hence, affects the system matrices in the system model (\ref{eq:system_plant}) and the output matrices in the output equations, such as (\ref{eq:output_I_B}), (\ref{eq:output_V_B}), and (\ref{eq:output_v_t}). This indicates that the evolution of system state and output variables can be controlled by adjusting the SSV.

As mentioned earlier, RBS design (a) in Fig. \ref{fig:RBS_designs} can be adapted to realize all other designs (b) to (f) along with their hybrid combinations. Therefore, the above system model formulated for design (a) is also applicable to designs (b) to (f) by selectively fixing the relevant switch states to zero in their SSVs. More importantly, the unified modeling framework introduced in this study can be deployed to model any RBS of interest, providing a solid foundation for the simulation, analysis, control, and optimization of RBSs. 

\subsection{Model validation by simulation and experimental tests}\label{subsec:model-verification}    

To evaluate the simulation accuracy of the proposed RBS modeling framework during operation and dynamic reconfiguration, a three-cell RBS was designed based on the structure in Fig. \ref{fig:RBS_designs} (d), consisting of three cells and seven switches indexed from 1 to 7. The RBS was discharged with a constant load power of 8 W. During the discharging process, the battery system was sequentially reconfigured every 20 seconds to each configuration described by the SSV in Table \ref{tab:SSV_label}. Those configurations leading to short circuits or open circuits are ignored.
\vspace{-5pt}
\begin{table}
	\centering
	\caption{SSVs for 12 configurations of the RBS without incurring battery faults.}
    \vspace{-5pt}
	\label{tab:SSV_label}
	\begin{tabular}{c c|c c} \hline \hline 
		Config. index&  SSV &  Config. index&  SSV\\ \hline 
		1 & [1,0,0,0,0,1,0] & 7 & [1,0,0,1,0,1,1]\\ 
		2 & [1,0,0,1,0,0,1]& 8 & [1,0,1,1,0,1,1]\\ 
		3 & [1,0,0,0,1,0,1]& 9 & [0,1,0,0,1,0,1]\\ 
		4 & [0,1,0,1,0,0,1]& 10 & [0,1,0,0,0,1,0]\\ 
		5 & [0,1,0,1,0,1,1]& 11 & [1,0,1,0,0,0,0]\\ 
		6 & [1,0,1,1,0,0,1]& 12 & [1,0,1,0,0,1,0]\\ \hline\hline 
	\end{tabular}
\end{table}
\vspace{-5pt}

To simulate the system operation and dynamic reconfiguration using the developed RBS model, we need to identify all parameter values of battery cells, switches, and wires. For each deployed EVE ICR 18650 battery cell, the ECM parameters were identified through Hybrid Pulse Power Characterization (HPPC) test data \cite{shh1999pngv} based on trust-region-reflective method, resulting in mean absolute errors (MAEs) of 0.8 mV, 1.0 mV, and 1.1 mV for the three cells. 
For each switch, two N-channel MOSFETs are connected in series to enable bidirectional current blocking. When using (\ref{eq:rds}) to calculate each switch's resistance, $R^m_{\text{ds(on)},n}=4~\text{m}\Omega$ according to the datasheet, $R_{\text{wire},n}^{m}=4~\text{m}\Omega$ based on the measurement, and $R^m_{\text{ds(off)},n}$ is approximately $2~\text{M}\Omega$ estimated from the drain current and drain-source voltage at zero gate voltage and room temperature.
    
\begin{figure}[ht]
        \centering
        \includegraphics[width=\columnwidth]{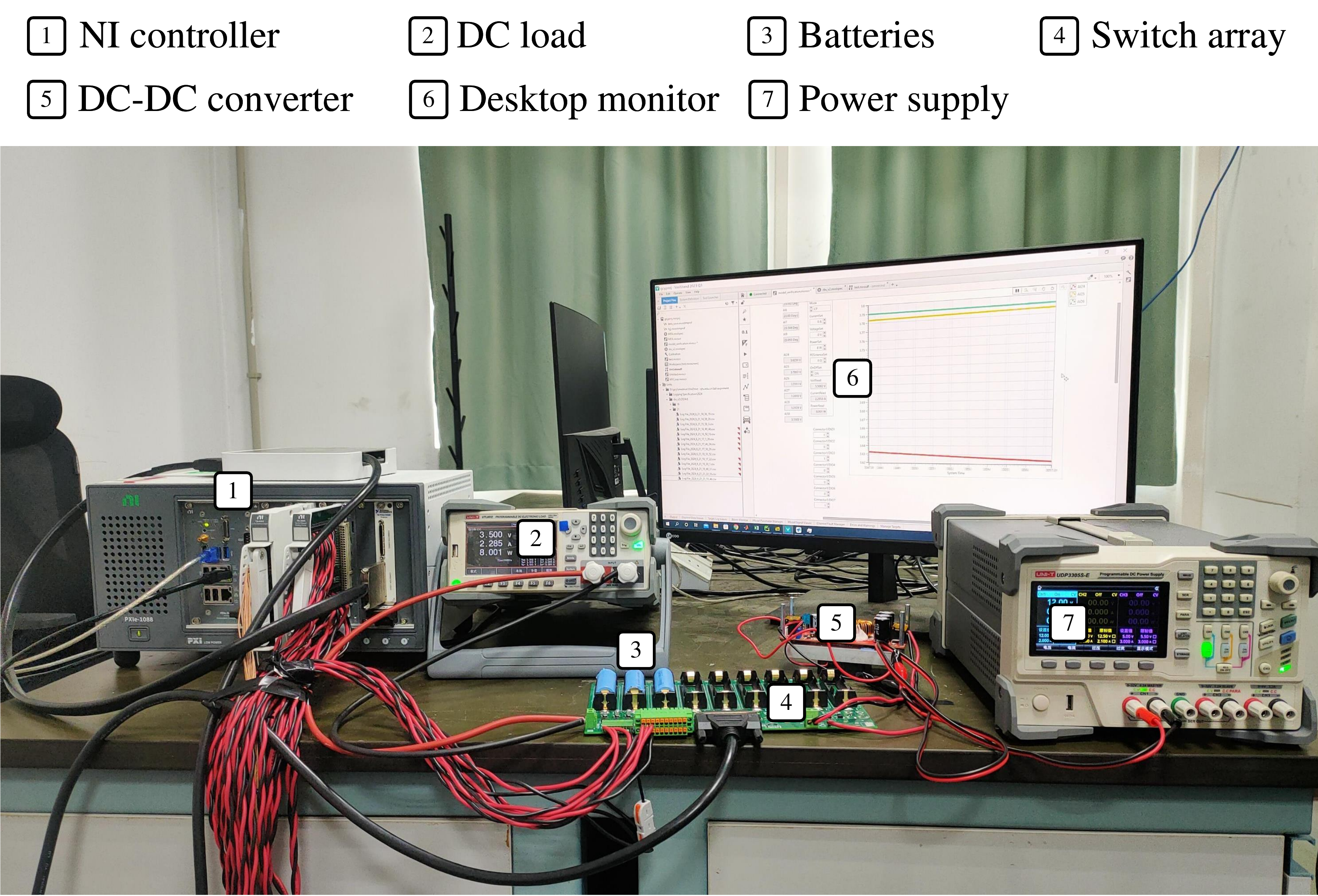}
        \caption{Illustration of the testbed for RBS experiments.}
        \label{fig:testbed}
    \end{figure}

To conduct the discharge test of the RBS, a testbed was set up, as shown in Fig. \ref{fig:testbed}.
Particularly, a drive circuit was designed for the MOSFET switch, and they were integrated on a printed circuit board (PCB) powered through a DC-DC converter according to the RBS structure design in Fig. \ref{fig:RBS_designs} (d). The RBS was discharged with a constant power by a direct current (DC) load. A data acquisition module is used to collect the voltage and current measurement data of battery cells at a sampling frequency of 10 Hz. A desktop computer and a National Instruments (NI) controller are deployed to generate control signals and send them to the drive circuits of switches. 
\vspace{-5pt}
\vspace{-5pt}

\begin{figure}[ht]
        \centering
        \includegraphics[width=\columnwidth]{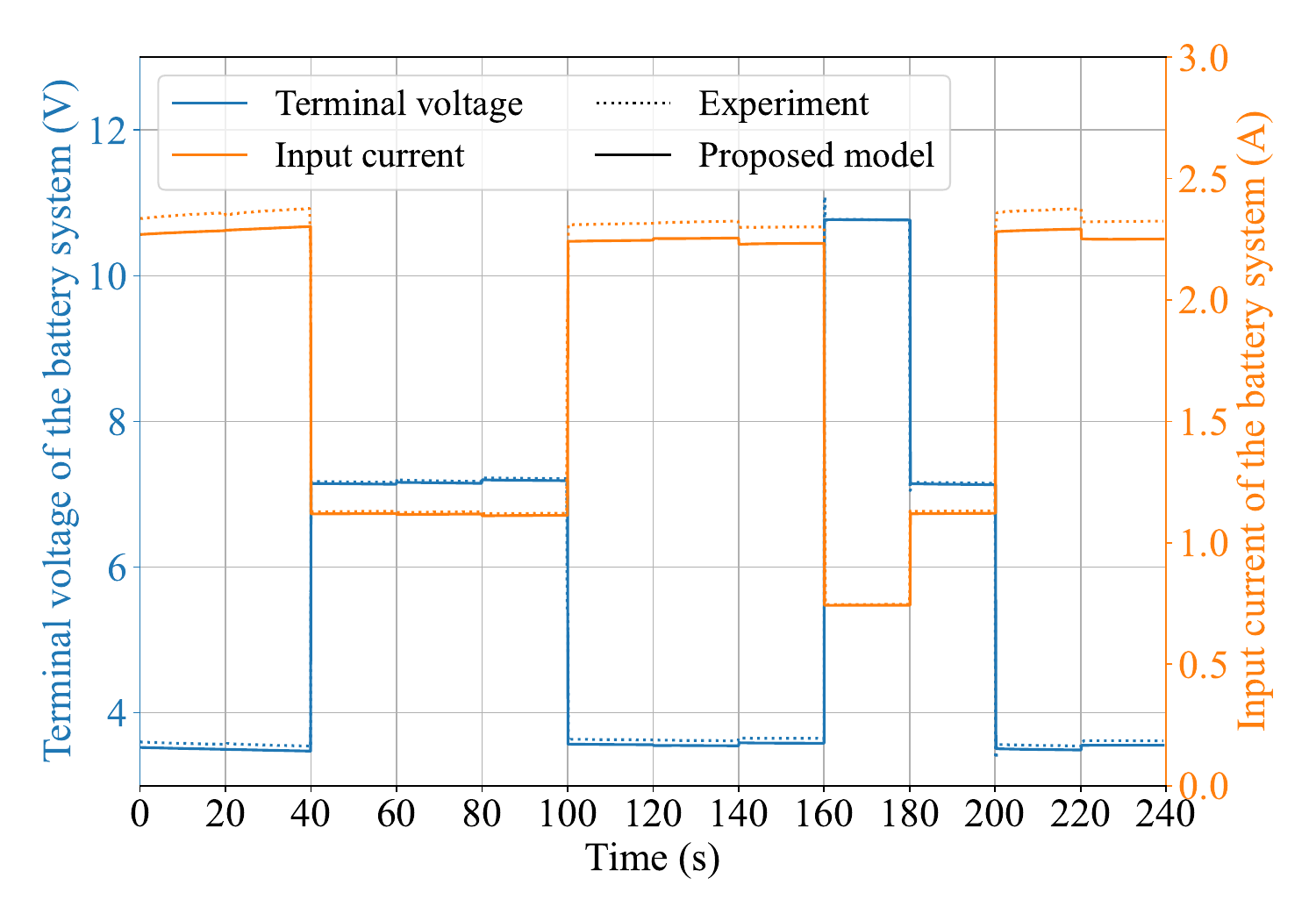}
        \vspace{-25pt}
        \caption{Comparison of the battery system's terminal voltage and input current obtained from model-based simulation with experimental measurement.}
        \label{fig:tv}
    \end{figure}

In Fig. \ref{fig:tv}, when the battery system is sequentially reconfigured from Config. 1 to Config. 12 every 20 seconds, the simulated evolution of the system's terminal voltage and input current, based on (\ref{eq:output_v_t}) and (\ref{eq:power2current}) in the proposed unified RBS model, are compared with experimental measurement. The simulation MAEs are 52 mV and 46 mA, respectively. Due to system reconfiguration, both terminal voltage and input current change accordingly despite the constant discharge power. 
\vspace{-5pt}

\begin{figure}[ht]
        \centering
        \includegraphics[width=\columnwidth]{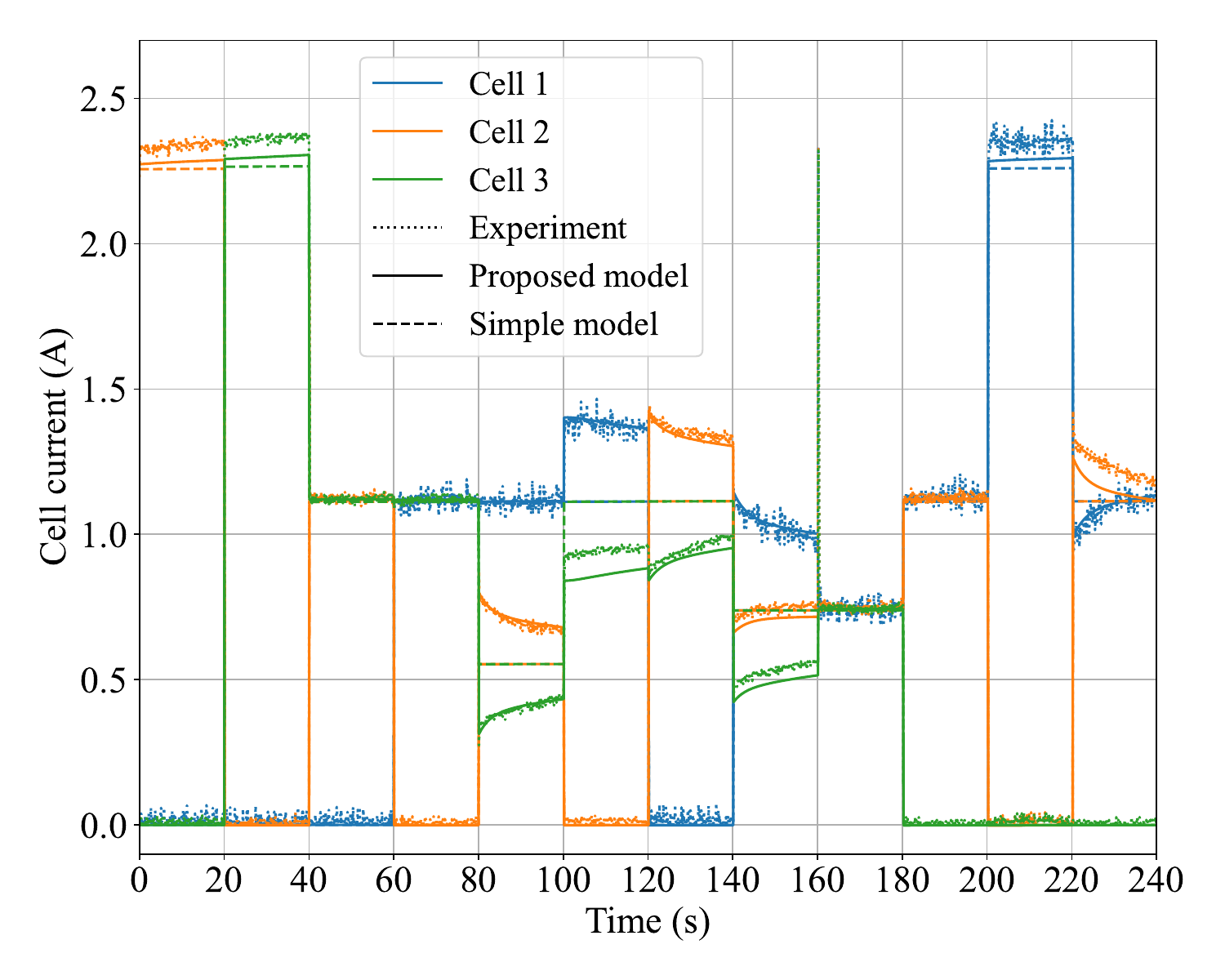}
        \vspace{-25pt}
        \caption{Comparison of the cell current evolution obtained by model-based simulation with experimental results.}
        \label{fig:cell_current}
    \end{figure}

\begin{figure}[ht]
        \centering
        \includegraphics[width=\columnwidth]{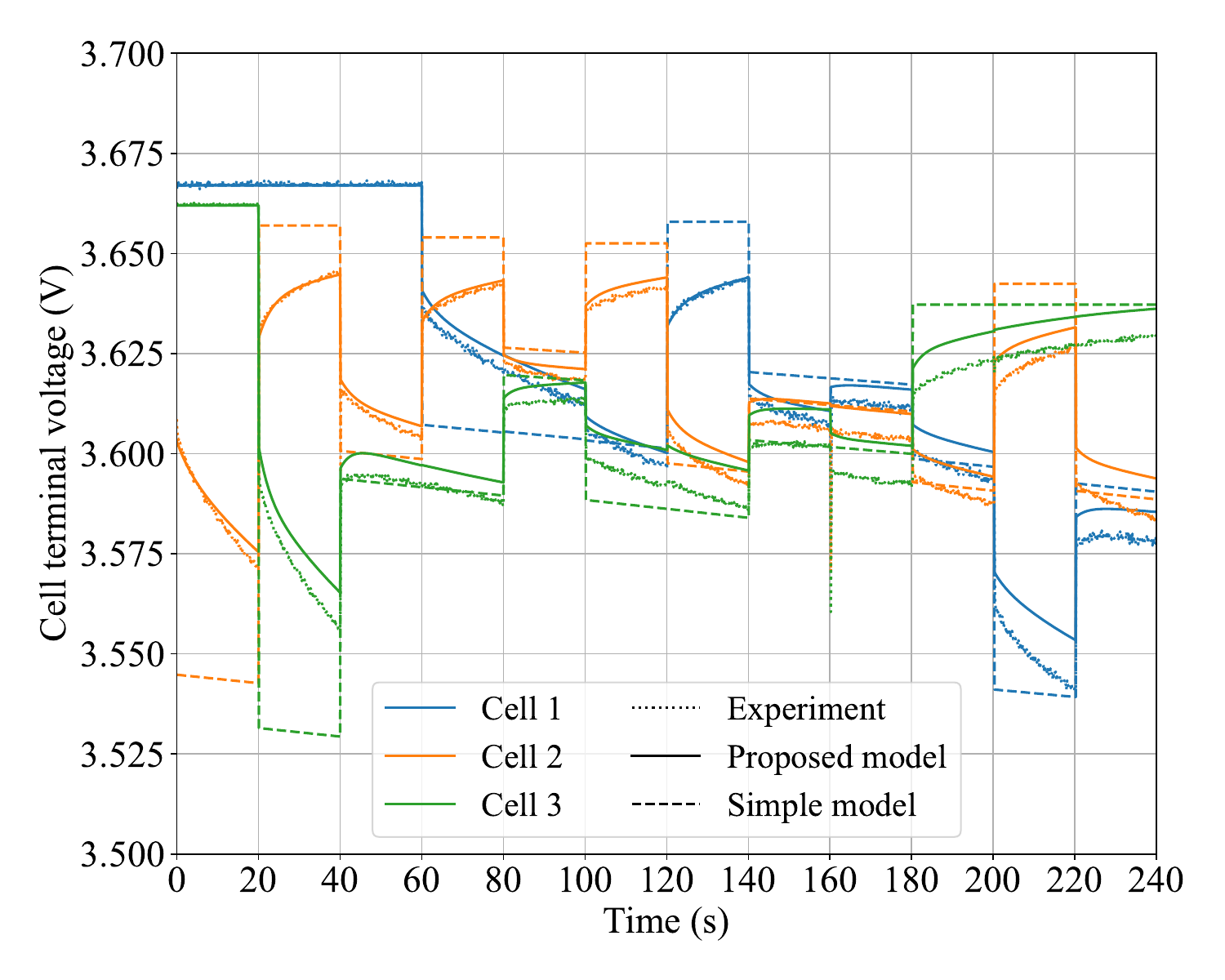}
        \vspace{-25pt}
        \caption{Comparison of the cell terminal voltage evolution obtained by model-based simulation with experimental results.}
        \label{fig:cell_voltage}
    \end{figure}

In addition to the system-level simulation, the simulated cell-level currents and voltages obtained from a simple model and the proposed model are compared with experimental results in Figures \ref{fig:cell_current} and \ref{fig:cell_voltage}, respectively. The simple model shares the same battery parameter values with the proposed model, but the total current is always assumed to be equally divided among parallel branches by neglecting the resistances of switches and wires in the simple model. According to the simulation MAEs for the proposed and simple models compared in Table \ref{tab:error_table}, we can see that the cell-level simulation results from the proposed model have current MAEs $\leq 20$ mA and voltage MAEs $\leq 7$ mV, much lower than those from the commonly used simple model assuming even parallel current distribution. 
The low simulation MAEs of the proposed model are mainly attributed to its capability of accurately characterizing the time-varying uneven current distribution among parallel cells during time intervals such as 80$\sim$160 seconds and 220$\sim$240 seconds in Fig. \ref{fig:cell_current}, corresponding to Config. 5 to Config. 8 and Config. 12 in Table \ref{tab:SSV_label}, respectively. In such configurations with parallel connections, capturing the dynamic current distribution is crucial for improving state estimation accuracy and keeping safe system operation. 
\vspace{-5pt}
\begin{table}
\centering
\caption{MAE of proposed and simple model.}
\label{tab:error_table}
\begin{tabular}{ c|c c c|c c c}\hline\hline
 & \multicolumn{3}{c|}{Proposed model}& \multicolumn{3}{c}{Simple model}\\
 \hline
 Cell index&  1& 2&3& 1& 2&3\\
Cell current (mA)& 19& 20&20& 69& 60&74\\
Terminal voltage of cell (mV)& 3.8& 5.1& 6.7& 7.7&  11.7&8.8\\
\hline\hline
\end{tabular}
\end{table}

In general, the simulated voltages and currents at system and cell levels based on the proposed unified RBS model can align well with the experimental measurement, except for some transient pulses. When reconfiguring the system from Config. 8 to Config. 9 at 160 s, some current or voltage pulses can be observed in Fig. \ref{fig:tv}, Fig. \ref{fig:cell_current}, and Fig. \ref{fig:cell_voltage} due to suffering from a short circuit fault during the switch operations. Fortunately, by considering the delay of control signals and appropriately designing the sequence of switch operations, the duration of such short circuit faults can be well restricted to minimize their impact on operation safety and performance control.  

\section{Construction of feasible solution space}\label{sec:construction-of-narrowed-search-space}
After addressing the critical challenge of formulating a unified RBS model in Section \ref{sec:unified-modeling-framework}, we are still confronted with the issue of low search efficiency when solving the formulated optimal control problems since the search space $\mathcal{B}^{N_S}$ grows exponentially with the increasing number of switches $N_S$.
In this section, we will develop a framework to narrow the search space for high efficiency as well as safe connections. At first, the mapping relation between a system configuration and a system SSV will be established. Then, an algorithm is developed to enumerate all feasible system configurations. Finally, based on the established mapping relation, only those feasible system configurations will be converted to feasible SSVs to construct a feasible search space $\mathcal{B}_f^{N_S}$. By tailoring the search scope from all system configurations to only those feasible ones, the size of the constructed feasible search space is expected to be much smaller than that of the complete search space. Since RBS design (a) in Fig. \ref{fig:RBS_designs} can be turned into other designs (b)-(f) by always connecting certain switches, it is taken as an example for elaboration on how to narrow the search space. 
\vspace{-7pt}
\subsection{Mapping a system configuration with a system SSV}
In a battery system, all battery cells connected in series (or parallel) in a local area form an internally series-connected (or parallel-connected) battery module, referred to as a \textit{series} (or \textit{parallel}) \textit{module}, denoted by $M_s$ (or $M_p$). 
If one single cell does not fall into any series or parallel module, it is regarded as a $M_s$ (or $M_p$) if cells are connected in series (or parallel) in its neighboring module.
Similarly, all local battery modules can construct a \textit{series} or \textit{parallel} \textit{battery pack}. 
Through analyzing all feasible configurations of RBS design (a), we can see that at most one battery pack can be constructed.
If battery cells are selected to form only series modules, then all series modules can constitute a parallel pack. This is referred to as a \textit{series-then-parallel configuration}, denoted by $C_{sp}$. On the other hand, battery cells can be selected to form some series and parallel modules, and all modules can then be connected in series to constitute a series pack. Such a configuration is referred to a \textit{parallel-then-series configuration}, denoted by $C_{ps}$. 

In such a hierarchical description of a battery pack, each battery cell belongs to one unique associated module, and further one unique associated pack.  
Once a battery cell's position and connection pattern (P\&CP), e.g., series or parallel, within its associated module as well as its associate module's P\&CP within its associated pack are specified, the states of all associated switches with the cell, referred to as a \textit{cell SSV}, will be determined. For instance, for the circled cell in RBS design (a) in Fig. \ref{fig:RBS_designs}, its cell SSV consists of the states of switches 1 to 5. All cell SSVs further compose the RBS's SSV to describe the system's configuration. 
The P\&CPs of a battery cell and its associated module are referred to as \textit{cell-module P\&CPs}. 

\begin{figure}[ht]
 \vspace{-10pt}
        \centering
        \includegraphics[width=\columnwidth]{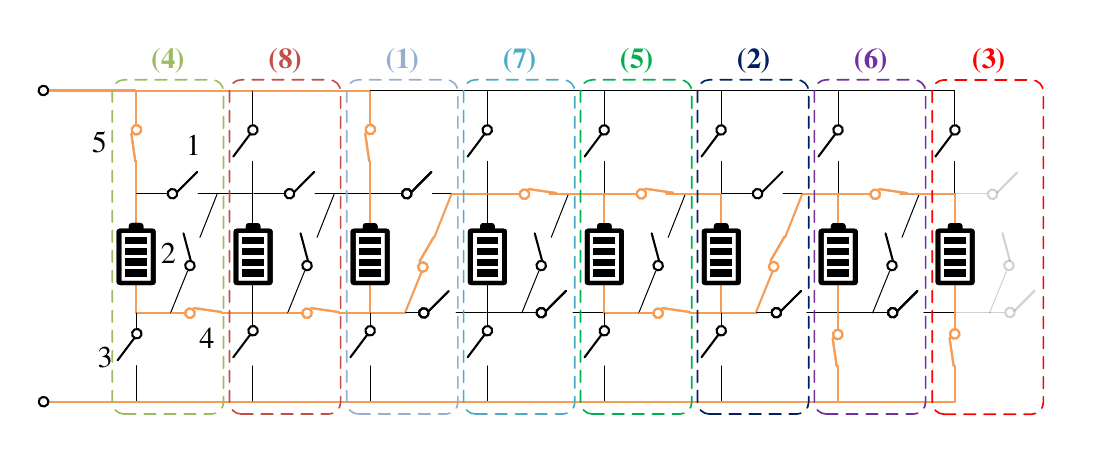}
        \vspace{-20pt}
        \caption{Examples of all eight possible cell connection patterns in RBS design (a) in Fig. \ref{fig:RBS_designs}. }
        \label{fig:position}
    \end{figure}

All possible cell-module P\&CPs are summarized in Table \ref{tab:switch_modes}, and the corresponding cell SSVs are listed and indexed. These cell SSVs are illustrated in Fig. \ref{fig:position} as well. 
As shown in Table \ref{tab:switch_modes}, multiple cell-module P\&CPs can be described by the same cell SSV, e.g., cell SSV (1). Moreover, if multiple cell SSVs lead to the same cell-module P\&CPs, only one of them is reserved in Table \ref{tab:switch_modes} . For instance, a cell can get connected to the next cell in parallel to construct a parallel module by connecting switches 4 and 5 (i.e., cell SSV (4)), switches 1 and 4 (i.e., cell SSV (5)), or switches 1 and 3 (i.e., cell SSV (6)). Among these three cell SSVs, the first and third ones can facilitate bypassing cells but they need the top or bottom bus lines for linking the two battery system terminals. Thus, cell SSVs (4) and (6) are assigned to construct the $M_p$'s at the beginning and end of a $C_{ps}$, respectively, and cell SSV (5) is used for constructing a $M_p$ at the middle of a $C_{ps}$.   
\begin{table*}[ht]
        \caption{Cell SSVs for all possible cell-module P\&CPs in RBS design (a) in Fig. \ref{fig:RBS_designs}.}
        \vspace{-5pt}
        \label{tab:switch_modes}
        \centering
        \scalebox{1}{
            \begin{tabular}{l c c}
                \hline\hline
                 \textbf{Cell-module P\&CPs} & \textbf{Cell SSV} & \textbf{Index}\\ \hline
                 The first cell in a $M_s$ in a $C_{sp}$, a $M_s$ at the beginning of a $C_{ps}$, or the last cell in a $M_p$ at the beginning of a $C_{ps}$.& [0, 1, 0, 0, 1] & (1)\\ 
                Middle cells in a $M_s$ in a $C_{sp}$, cells in a $M_s$ in a $C_{ps}$, or the last cell in a $M_p$ at the middle of a $C_{ps}$.& [0, 1, 0, 0, 0] & (2)\\ 
                The last cell in a $M_s$ at the end of a $C_{ps}$, a $M_s$ in a $C_{sp}$, or a $M_p$ at the end of a $C_{ps}$.& [0, 0, 1, 0, 0]  & (3)\\ 
                The first cell or middle cells in a $M_p$ at the beginning of a $C_{ps}$.  & [0, 0, 0, 1, 1] & (4)\\ 
                The first cell or middle cells in a $M_p$ at the middle of a $C_{ps}$. & [1, 0, 0, 1, 0] & (5)\\ 
                The first cell or middle cells in a $M_p$ at the end of a $C_{ps}$. & [1, 0, 1, 0, 0] & (6)\\ 
                The bypassed cell in a $M_s$ in a $C_{ps}$, or a $M_p$ at the middle or end of a $C_{ps}$.  & [1, 0, 0, 0, 0] & (7)\\ 
                The bypassed cell in a $M_p$ at the beginning of a $C_{ps}$, or a $M_s$ in a $C_{sp}$ .  & [0, 0, 0, 1, 0] & (8)\\ \hline\hline
            \end{tabular}}
\end{table*}
\vspace{-5pt}

Thus, given any configuration of the RBS design (a), we can first identify each cell's cell-module P\&CPs and then look up its corresponding cell SSV based on Table \ref{tab:switch_modes}. All the obtained cell SSVs are finally used to construct the system SSV. Next, in order to formulate the feasible search space of system SSVs we need to identify all those feasible system configurations. 
\vspace{-7pt}
\subsection{Identifying all feasible system configurations}
Among all configurations of an $N_B$-cell RBS, there exist a majority of infeasible system configurations resulting in short circuit faults, unacceptable open circuits, or undesired system voltages. 
Thus, algorithms are developed to identify the remaining feasible configurations. 

To construct a feasible system configuration meeting the required system voltage, it is necessary to specify the number of series-connected battery cells along the connection path between the system's positive and negative terminals, which is referred to as the \textit{normalized system voltage} and denoted by $\tilde v_t$.
For an RBS with $N_B$ cells, the integer-value normalized system voltage $\tilde v_t\in\{1,\ldots,N_B\}$. Given the required system voltage $v_t^{req}\in[v_t^{min}, v_t^{max}]$, the required normalized system voltage $\tilde v_t^{req}$ needs to be specified accordingly to facilitate the configuration construction, i.e., $\tilde v_t^{req}\in\{\tilde v_t^{min}, \dots, \tilde v_t^{max}\}$. 

The lower and upper limits of the normalized system voltage, $\tilde v_t^{min}$ and $\tilde v_t^{max}$, can be generated from the original system voltage limits, $v_t^{min}$ and $v_t^{max}$, in different ways. According to (\ref{eq:v^B_n}), each battery cell's terminal voltage is not constant but varies with multiple factors, e.g., its OCV and current. Denote the nominal voltage and the maximum and minimum allowable terminal voltages of a battery cell by $v^B_{nom}$, $v^B_{max}$, and $v^B_{min}$, respectively. To cover various operating scenarios when constructing the feasible configurations, $\tilde v_t^{min}$ and $\tilde v_t^{max}$ can be roughly derived based on the battery cell's nominal voltage, i.e., $\tilde v_t^{min}=\lfloor \frac{v_t^{min}}{v^B_{nom}} \rfloor$ and $\tilde v_t^{max}=\lceil \frac{v_t^{max}}{v^B_{nom}} \rceil$. To accommodate all potentially feasible configurations under certain operating scenarios, the selection range of $\tilde v_t$ can be further extended to, e.g., $\{\lfloor \frac{v_t^{min}}{v^B_{min}} \rfloor,\ldots, \lceil \frac{v_t^{max}}{v^B_{max}} \rceil\}$. Through these ways, some configurations might fail to meet the required system voltage under certain operating scenarios, but they can be further screened by setting relevant operating constraints when formulating control or optimization problems.

In a series-then-parallel configuration $C_{sp}$, 
all series-connected cells along with those bypassed cells in between in each series module $M_s$ are neighboring cells. 
If the $C_{sp}$ needs to meet the required normalized system voltage $\tilde v_t^{req}$, each $M_s$ in the $C_{sp}$ must have $\tilde v_t^{req}$ series-connected cells. For RBS design (a), any cell in a $C_{sp}$ can be bypassed. 
In a $C_{sp}$ of an $N_B$-cell RBS with $N_{M_s}$ parallel modules, $N_{M_s}\in[2,\lfloor \frac{N_B}{\tilde v_t^{req}} \rfloor]$, denote the $j$-th series module by $M_{s,j}$, $1\leq j \leq N_{M_s}$, the $k$-th series-connected cell in $M_{s,j}$ by $c_{j,k}$, $1\leq k \leq \tilde v_t^{req}$, and the index of cell $c_{j,k}$ among all cells in the RBS by $l_{j,k}\in[1,N_B]$. Then, a $C_{sp}$ can be constructed in terms of cell indices as follows by sequentially selecting $\tilde v_t^{req}$ battery cells for each $M_s$ while skipping those bypassed cells. As a result, a total of $\sum\limits_{N_{M_s}=2}^{\lfloor \frac{N_B}{\tilde v_t^{req}} \rfloor} \binom{N_{M_s}\tilde v_t^{req}}{N_B}$ feasible $C_{sp}$'s are obtained, in which $\binom{m\tilde v_t^{req}}{N_B}$ calculates a binomial coefficient.
\begin{align*}
&C_{sp} = (M_{s,1}, M_{s,2}, \dots, M_{s,N_{M_s}}), \\
&M_{s,j} = (l_{j,1}, l_{j,2}, \dots, l_{j,\tilde v_t^{req}}), \\
\end{align*}
\begin{align*}
& 1\leq l_{j,1} < l_{j,2}<\ldots<l_{j,\tilde v_t^{req}}<l_{j+1,1}\leq N_B,\\
&N_{M_s},j \in \mathbb{Z},~ 1<N_{M_s}\leq \frac{N_B}{\tilde v_t^{req}}, ~1\leq j\leq N_{M_s}.
\end{align*}

Different from the above series-then-parallel configurations allowing only series modules with the same number of cells and arbitrary cell bypassing, a parallel-then-series configuration $C_{ps}$ allows both series and parallel modules, and the number of cells in each module can differ from each other. Moreover, in a $C_{ps}$ of RBS design (a), all cells can be bypassed except those in a $M_p$ at the middle of the $C_{ps}$. 
Consequently, it becomes more difficult to identify all feasible $C_{ps}$'s.
To address this challenge, Algorithm \ref{alg1} is developed to identify all feasible $C_{ps}$'s using dynamic programming (DP), which creates sets iteratively, leveraging information from previously established sets to optimize brute-force enumeration. By storing and reusing intermediate results to avoid redundant calculations, the search time can be significantly reduced.
\begin{algorithm*}[ht]
\caption{Algorithm for enumerating all feasible parallel-then-series configurations}
\label{alg1}
\begin{algorithmic}[1]
\State Define five cell position flags: in the $M_p$ at the beginning of a $C_{ps}$ ($P_{\text{bgn}}$), in any $M_p$ at the middle of a $C_{ps}$ ($P_{\text{mid}}$), in the $M_p$ at the end of a $C_{ps}$ ($P_{\text{end}}$), in any $M_s$ ($S$), and bypassed ($B$). Then, represent a $C_{ps}$ by a vector consisting of all cells' flags, denoted by $V_{cp}$.
\State Initialize $VL$ as a $ N_B \times N_B \times 6$ list of cell flag vectors, and let $VL[i][j][:]$ represent the list of $V_{cp}$'s with a total of $i$ cells and reaching the normalized system voltage $j$. If the last non-bypassed cell's position flag in a $V_{cp}$ with $i$-cell and meeting $j$-level is $P_{\text{bgn}}$, $S$, or $P_{\text{end}}$, the $V_{cp}$ will be placed into the sub-list $VL[i][j][0]$, $ VL[i][j][1] $, or $ VL[i][j][2]$, respectively. The other sub-lists store transitional $V_{cp}$'s established only for augmentation. If all cells in the last module are labeled by \(P_{\text{mid}}\), one single cell labeled by \(P_{\text{bgn}}\) constitutes the first module, or one single cell labeled by \(P_{\text{end}}\) constitutes the last module, the corresponding transitional $V_{cp}$ will be placed in $VL[i][j][3]$, $ VL[i][j][4] $, or $ VL[i][j][5]$, respectively.
\State Initialize $VL[1][1][1] \gets [[S]]$
\State Initialize $VL[1][1][4] \gets [[P_{\text{bgn}}]]$
\For{$i \gets 2$ \textbf{to} $N_B$}
    \For{$j \gets 1$ \textbf{to} $i$}
        \If{$j = 1$}
            \State Initialize $all\_bypass \gets [B] \times (i - 1)$ \Comment{Note: $all\_bypass$ is a $V_{cp}$ with all $i-1$ entries equal to $B$.} 
            \State $VL[i][j][1].append(all\_bypass.append(S))$ 
            \State $VL[i][j][4].append(all\_bypass.append(P_{\text{bgn}}))$
        \EndIf

        \State \parbox[t]{\dimexpr\linewidth-\algorithmicindent *2}{$VL[i][j][0].extend([h.append(B) \mid h \in VL[i - 1][j][0]])$ \Comment{Note: $h$ represents any $V_{cp}$ extracted from a sub-list.}} 
        \State \parbox[t]{\dimexpr\linewidth-\algorithmicindent *2}{$VL[i][j][0].extend([h.append(P_{\text{bgn}}) \mid h \in VL[i - 1][j][0]  ~\text{or}~ VL[i - 1][j][4]])$}
        \State \parbox[t]{\dimexpr\linewidth-\algorithmicindent *2}{$VL[i][j][1].extend([h.append(B) \mid h \in VL[i - 1][j][1]])$}
        \State \parbox[t]{\dimexpr\linewidth-\algorithmicindent *2}{$VL[i][j][1].extend([h.append(S) \mid h \in VL[i - 1][j - 1][0], ~VL[i - 1][j - 1][1], ~\text{or}~ VL[i - 1][j - 1][3]])$}
        \State \parbox[t]{\dimexpr\linewidth-\algorithmicindent *2}{$VL[i][j][2].extend([h.append(B) \mid h \in VL[i - 1][j][2]])$}
        \State \parbox[t]{\dimexpr\linewidth-\algorithmicindent *2}{$VL[i][j][2].extend([h.append(P_{\text{end}}) \mid h \in VL[i - 1][j][2] ~\text{or}~ VL[i - 1][j][5]])$}
        \State \parbox[t]{\dimexpr\linewidth-\algorithmicindent *2}{$VL[i][j][3].extend([h.append(B) \mid h \in VL[i - 1][j][3]])$}
        \For{$n\_mp \gets 2$ \textbf{to} $i - 1$} \Comment{Note: $n\_mp$ is the number of cells in the $M_p$ at the middle of a $C_{ps}$.}
            \State \parbox[t]{\dimexpr\linewidth-\algorithmicindent *3}{$VL[i][j][3].extend([h.append(P_{\text{mid}}) \times n\_mp \mid h \in VL[i - n\_mp][j - 1][0],~ VL[i - n\_mp][j - 1][1] ~\text{or}~ VL[i - n\_mp][j - 1][3]])$}
        \EndFor
        \State \parbox[t]{\dimexpr\linewidth-\algorithmicindent *2}{$VL[i][j][4].extend([h.append(B) \mid h \in VL[i - 1][j][4]])$}
        \State \parbox[t]{\dimexpr\linewidth-\algorithmicindent *2}{$VL[i][j][5].extend([h.append(B) \mid h \in VL[i - 1][j][5]])$}
        \State \parbox[t]{\dimexpr\linewidth-\algorithmicindent *2}{$VL[i][j][5].extend([h.append(P_{\text{end}}) \mid h \in VL[i - 1][j - 1][0], ~VL[i - 1][j - 1][1] ~\text{or}~ VL[i - 1][j - 1][3]])$}
    \EndFor
\EndFor
\State \textbf{return} $VL[N_B][\tilde v_t^{min}:\tilde v_t^{max}][0:2]$
\end{algorithmic}
\end{algorithm*}

As shown in Line 1 in Algorithm \ref{alg1}, the cell-module P\&CPs listed in Table \ref{tab:switch_modes} are further aggregated, and five types of position flags of battery cells are defined, including the cell bypassing. Then, a $C_{ps}$ can be described by a vector consisting of all cells' position flags, denoted by $V_{cp}$. To accommodate all $V_{cp}$'s, a three-dimensional list $VL$ is constructed in Line 2. The first dimension $i$ describes the number of cells in the system. The second dimension $j$ specifies the required system voltage level. Moreover, six types of sub-lists are constructed based on the third dimension. Among them, sub-lists $ VL[i][j][0] $, $ VL[i][j][1] $, and $ VL[i][j][2]$ store $V_{cp}$'s corresponding to $C_{ps}$'s consisting of only one $M_p$, having an $M_s$ at the end, or having an $M_p$ at the end, respectively. These sub-lists are separated since they need to be constructed in different ways.   
Sub-lists $ VL[i][j][3] $, $ VL[i][j][4] $, and $ VL[i][j][5]$ store $V_{cp}$'s only used for transition, and, hence, they are not returned eventually. 

Next, from Line 3 to Line 30, DP is deployed to iteratively search for the $V_{cp}$'s corresponding to all feasible $C_{ps}$'s. For single-cell subsystems, sub-lists $VL[1][1][1]$ and $VL[1][1][4]$ are initialized in Lines 3 and 4, respectively, in which the inner brackets denote a vector and the outer brackets denote a vector list. For any multi-cell subsystem with $i>1$, sub-lists $VL[i][1][1]$ and $VL[i][1][4]$ need to be initialized in Lines 9 and 10, respectively, based on the $all\-bypass$ vector defined in Line 8. By appending a new item with value $x$ to the end of an array $y$ through the $append$ operation, an augmented array $y.append(x)$ can be obtained. 
For example, in Line 9 a cell position flag of $S$ is appended to the end of the vector $all\-bypass$ to get an augmented vector $all\-bypass.append(S)$, which is then appended to the end of the vector list $VL[i][j][1]$. 

Then, for each normalized system voltage $j\in[1, i]$ of the $i$-cell subsystem, six types of sub-lists are constructed in Lines 12 to 24 through the $extend$ operation. Given two vector lists $x$ and $y$, $y.extend(x)$ represents an augmented vector list combining all vectors in both lists. In particular, the expression $h.append(P_{\text{mid}}) \times n_{mp}$ in Lines 20 describes an augmented vector by appending $n_{mp}$ $P_{\text{mid}}$ cell position flags to the original vector $h$, indicating that a $M_p$ composed of $n_{mp}$ cells is connected in series with the original system with $i-n_{mp}$ cells. All possible ways need to be considered to construct the $V_{cp}$'s for each type of sub-list. For instance, a $V_{cp}$ for the sub-list $VL[i][j][1]$ can be constructed by appending a flag $B$ to the end of any $V_{cp}$ in sub-list $VL[i-1][j][1]$ or a flag $S$ to the end of any $V_{cp}$ in sub-lists $VL[i-1][j-1][0]$, $VL[i-1][j-1][1]$, or $VL[i-1][j-1][3]$, as shown in Lines 14 or 15. Through such iterative $append$ and $extend$ operations, the feasible $C_{ps}$'s for the investigated $N_B$-cell system can be gradually approached. Finally, the first three types of sub-lists $VL[N_B][\tilde v_t^{min}:\tilde v_t^{max}][0:2]$ for the $N_B$-cell system meeting the required normalized system voltages $\tilde v_t^{req}\in[\tilde v_t^{min}, \tilde v_t^{max}]$ are returned for enumerating all feasible $C_{ps}$'s of the RBS.    

According to the mapping relation established in Table \ref{tab:switch_modes}, all the identified feasible $C_{sp}$'s and $C_{ps}$'s are translated to feasible SSVs, which compose the feasible search space for system SSVs $\mathcal{B}_f^{N_S}$.
Furthermore, since RBS designs (b)–(f) can be realized by removing certain switch branches in RBS design (a) as shown in Fig. \ref{fig:RBS_designs}, their feasible search spaces can be obtained based on design (a)'s feasible search space by excluding those SSVs assigning state value 1 to the already removed switch branches. 
\vspace{-5pt}
\subsection{Quantifying the narrowing effect of the proposed feasible search space}
For an $N_B$-cell RBS with design (a) as shown in Fig. \ref{fig:RBS_designs} (a), the total number of switches is $N_S=5N_B-3$ since each cell is associated with five switches except the last one that has only two switches. Then, the total number of possible SSVs in the complete search space $\mathcal{B}^{N_S}$ is $2^{N_S}=2^{5N_B-3}$. When constructing the feasible search space of SSVs, denoted by $\mathcal{B}_f^{N_S}$, both the system size and the required normalized system voltage need to be considered. Thus, denote the total number of feasible SSVs meeting the normalized system voltage $\tilde v_t$ in $\mathcal{B}_f^{N_S}$ by $N_{f}{(N_B, \tilde v_t)}$. Then, the total number of feasible SSVs across all normalized system voltages is $\sum_{\tilde v_t=1}^{N_B}N_{f}{(N_B,\tilde v_t)}$. 

To demonstrate the narrowing effect of the identified feasible search space, 
the number of feasible SSVs in the $\mathcal{B}_f^{N_S}$ is compared with the number of all possible SSVs in the $\mathcal{B}^{N_S}$ in Table \ref{tab:cc_counts}.
The sizes of $\mathcal{B}_f^{N_S}$ and $\mathcal{B}^{N_S}$, in terms of the number of SSVs, both grow as the system size $N_B$ increases, but the former is much smaller than the latter and their ratios get even smaller as $N_B$ increases. Taking the 10-cell system as an example, even though all normalized system voltages are allowed, the size ratio of $\mathcal{B}_f^{N_S}$ and $\mathcal{B}^{N_S}$ is very tiny, less than $2\times10^{-10}$. Thus, the search effort is expected to be significantly reduced by narrowing the search space to only feasible SSVs, potentially enabling the solution of optimal control problems for RBSs. A case study is provided to demonstrate the effectiveness of the narrowed search space.
\begin{table*}[ht]
        \caption{Comparison of the number of feasible SSVs  $N_f(N_B,\tilde v_t)$ for RBSs with various cell numbers and normalized system voltages.}
        \vspace{-5pt}
        \label{tab:cc_counts}
        \centering
        \resizebox{1\linewidth}{!}{
        \begin{tabular}{|c|c|c|c|c|c|c|c|c|c|c|}
            \hline
            $N_{f}{(N_B,\tilde v_t)}$     & $N_B=2$ & $N_B=3$ & $N_B=4$ & $N_B=5$  & $N_B=6$  & $N_B=7$  & $N_B=8$& $N_B=9$ & $N_B=10$\\
            \hline
            $\tilde v_t=1$ & 3 & 7 & 15 & 31 & 63 & 127 & 255 & 511 & 1023 \\ \hline
            $\tilde v_t=2$ & 1 & 5 & 18 & 54 & 144 & 356 & 839 & 1919 & 4307 \\ \hline
            $\tilde v_t=3$ & - & 1 & 7 & 30 & 103 & 310 & 853 & 2200 & 5410 \\ \hline
            $\tilde v_t=4$ & - & - & 1 & 9 & 47 & 187 & 631 & 1907 & 5327 \\ \hline
            $\tilde v_t=5$ & - & - & - & 1 & 11 & 68 & 312 & 1186 & 3959 \\ \hline
            $\tilde v_t=6$ & - & - & - & - & 1 & 13 & 93 & 485 & 2063 \\ \hline
            $\tilde v_t=7$ & - & - & - & - & - & 1 & 15 & 122 & 714 \\ \hline
            $\tilde v_t=8$ & - & - & - & - & - & - & 1 & 17 & 155 \\ \hline
            $\tilde v_t=9$ & - & - & - & - & - & - & - & 1 & 19 \\ \hline
            $\tilde v_t=10$ & - & - & - & - & - & - & - & - & 1 \\ 
            \hline
            No. of feasible SSVs = $\sum_{\tilde v_t=1}^{N_B}N_{f}{(N_B,\tilde v_t)}$ & 4 & 13 & 41 & 125 & 369 & 1062 & 2999 & 8348 & 22978 \\
            \hline
            No. of possible SSVs = $2^{5N_B-3}$ & 128 & $4.096 \times 10^3$ & $1.311 \times 10^5$ & $4.194 \times 10^6$ & $1.342 \times 10^8$ & $4.295 \times 10^9$ & $1.374 \times 10^{11}$ & $4.398 \times 10^{12}$ & $1.407 \times 10^{14}$ \\
            \hline
            \text{$\sum_{\tilde v_t=1}^{N_B}N_{f}{(N_B,\tilde v_t)}/2^{5N_B-3}$} & $3.125 \times 10^{-2}$ & $3.174 \times 10^{-3}$ & $3.128 \times 10^{-4}$ & $2.980 \times 10^{-5}$ & $2.749 \times 10^{-6}$ & $2.473 \times 10^{-7}$ & $2.182 \times 10^{-8}$ & $1.898 \times 10^{-9}$ & $1.633 \times 10^{-10}$ \\
            \hline
            \end{tabular}}
    \end{table*}
    \vspace{-5pt}
    \subsection{Case study for comparing the searching effects within the complete and feasible search spaces}\label{subsec:simulation-result}
    To compare the searching effects within the complete search space $\mathcal{B}^{N_S}$ and the narrowed search space $\mathcal{B}_f^{N_S}$, 
    a 10-cell RBS with design (d) in Fig. \ref{fig:RBS_designs} is taken as an example. The RBS is discharged with a constant power of 70 W for 20 decision steps, and each decision step lasts 20 seconds. For each battery cell, the SoC is constrained between 0.05 and 1, and the discharge current rate can reach up to 6 C (corresponding to 12 A). The normalized system voltage is restricted to at least four (i.e., at least four cells or modules need to be connected in series).
    The initial SoCs for the 10 cells have a total SoC imbalance $SoC_{ibl}^{tot}=0.30$, which is defined as follows.
    \begin{equation}
        \label{eq:objective}
        \begin{aligned}
            SoC_{ibl}^{tot} = \sum_{n=1}^{N_B}\text{{SoC}}_n - N_B  \min_{n=1}^{N_B}\text{{SoC}}_n.
        \end{aligned}
    \end{equation}
    In particular, $SoC_{ibl}^{tot}$ represents the total unused SoC when any cell SoC reaches the lower limit, necessitating the termination of system operation. 
    Then, an RBS optimal control problem is formulated for minimizing the $SoC_{ibl}^{tot}$ at the end of the last decision step. A commonly used heuristic method, Genetic Algorithm (GA) \cite{johnh.1992Adaptation}, is deployed to solve this problem, in which the population size of individuals in each generation, the number of generations, crossover probability, and mutation probability are set to 100, 220, 0.8, and 0.1, respectively. 
    
\begin{figure}[ht]
        \centering
        \includegraphics[width=0.8\columnwidth]{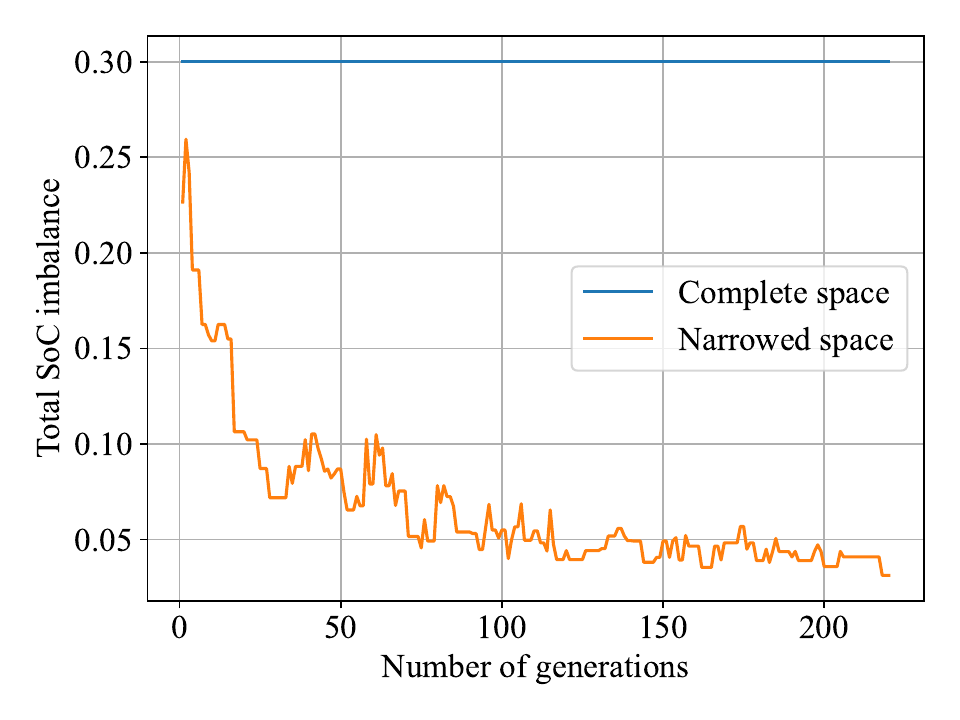}
        \vspace{-15pt}
        \caption{Comparison of optimization results based on the complete search space and the narrowed search space.}
        \label{ga}
    \end{figure}

\begin{figure}[ht]
        \centering
        \includegraphics[width=0.8\columnwidth]{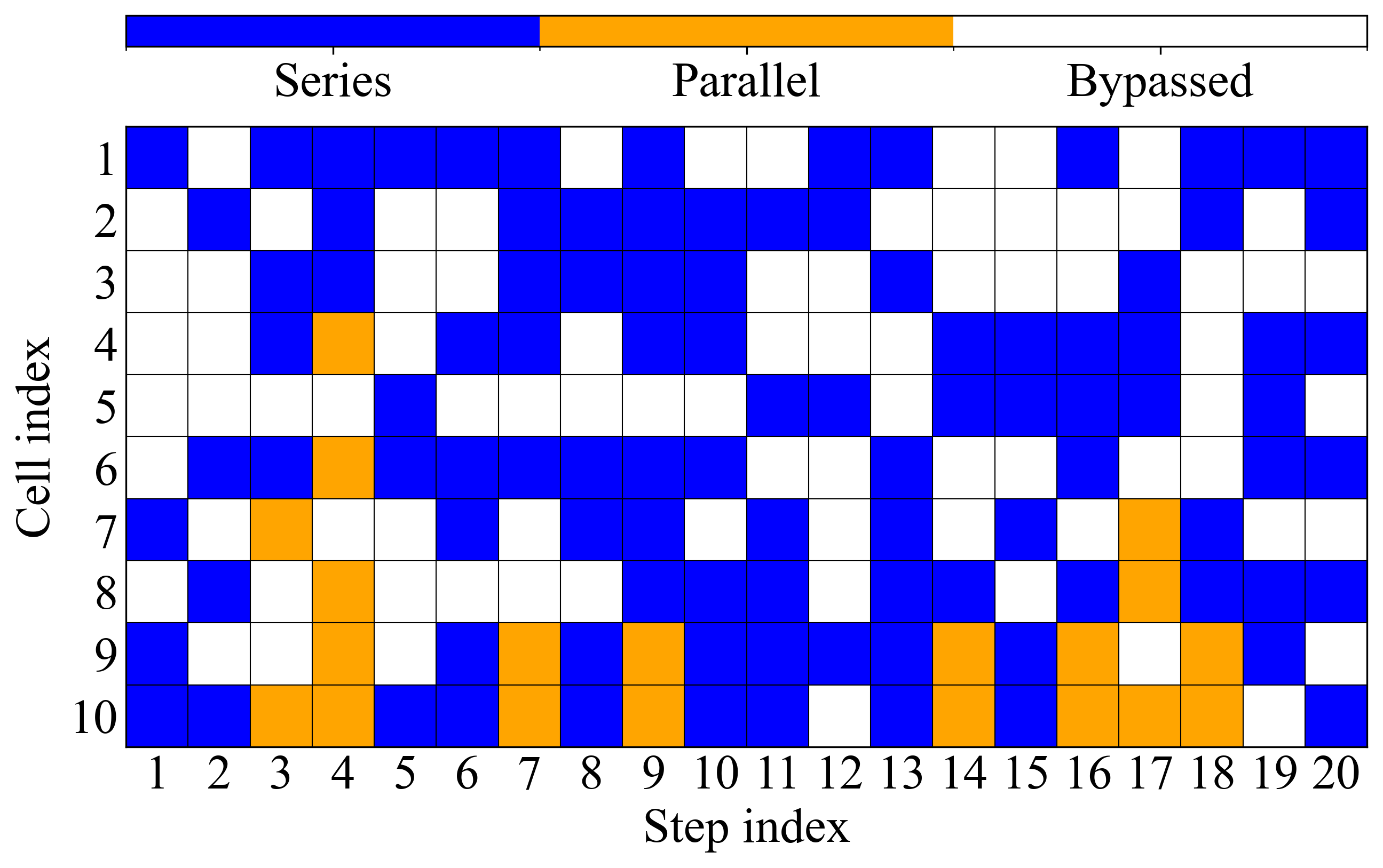}
        \vspace{-8pt}
        \caption{Optimized system configurations described by the cell connection patterns during all decision steps .}
        \label{sim_confs}
    \end{figure}

\begin{figure}[ht]
        \centering
        \includegraphics[width=0.9\columnwidth]{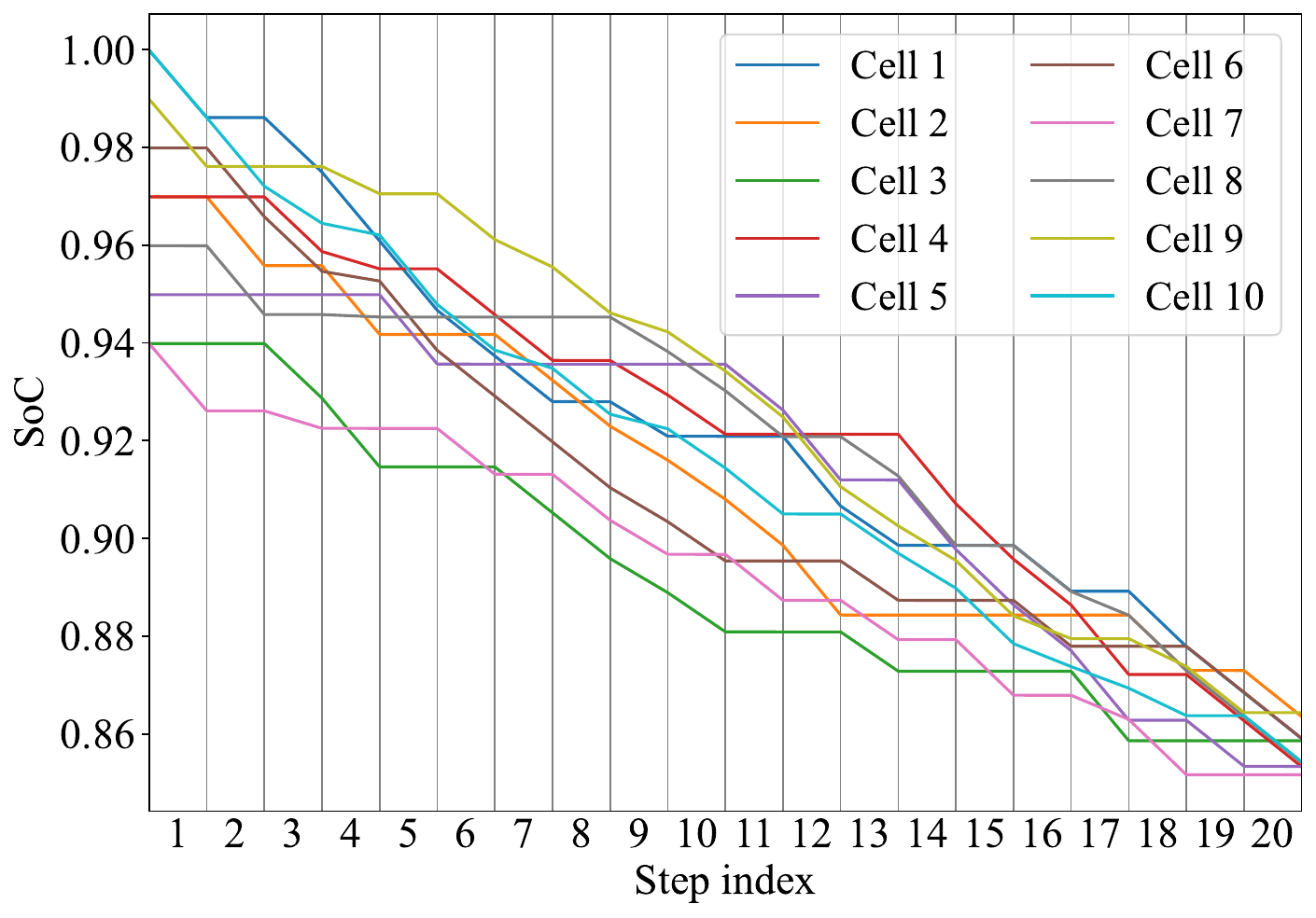}
        \vspace{-10pt}
        \caption{The cell SoC evolutions over decision steps under the optimized system configurations.}
        \label{sim_fittest}
    \end{figure}

    Both the complete search space and the feasible search space are tested when solving the problem, and the resulting evolutions of $SoC_{ibl}^{tot}$ over generations are compared in Fig. \ref{ga}. When searching within the complete search space, the $SoC_{ibl}^{tot}$ remains unchanged over successive generations, indicating that the problem remains unsolved. The primary reason is that the majority of SSVs in the complete search space are infeasible for each decision step, as shown in Table \ref{tab:cc_counts}. This makes it extremely challenging to select a feasible SSV for the initial decision step, let alone a series of consecutive feasible actions across all 20 decision steps. Indeed, it has been found that the SSVs generated for the first decision steps of all 100 individuals in each generation are infeasible. Consequently, for each generation, there is no need to evaluate the balancing performance of each individual across all decision steps, and the $SoC_{ibl}^{tot}$ is not improved over generations. 
    This challenge can be addressed if the search space is narrowed to only feasible SSVs. As a result, the GA using the feasible search space manages to reduce the $SoC_{ibl}^{tot}$ from 0.3 to 0.031 after 220 generations. Thus, the proposed feasible search space enables the convergence to the optimal solution. 
    Specifically, the optimized system configuration for each decision step is described in Fig. \ref{sim_confs} by battery cells' connection patterns within their associated modules, i.e., series, parallel, or bypassed. Moreover, the corresponding evolutions of cell SoCs over decision steps are presented in Fig. \ref{sim_fittest}.
    Clearly, the cell SoC imbalance gradually decreases over decision steps due to the optimized system configurations illustrated in Fig. \ref{sim_confs}. 
    \vspace{-5pt}
    
    \section{Conclusions}\label{sec:conclusion}
    This study aims to address two fundamental challenges in the optimal control of RBSs: constructing the system model expression during problem formulation and enhancing search efficiency during problem-solving.
    
    The proposed unified modeling framework based on the RBS's CSC eliminates the need to frequently reconstruct the system model during dynamic battery reconfiguration. This enables the formulation of optimal control problems for RBSs while significantly reducing computational time and storage requirements. Compared to experimental test results, the proposed RBS model demonstrates high accuracy in simulating dynamic current and voltage distributions within the system.
    
    Furthermore, when solving the formulated RBS optimal control problems, an algorithm is proposed to narrow the complete search space to only feasible solutions, thereby not only dramatically enhancing search efficiency but also filtering out all configurations that could result in battery faults or unacceptable terminal voltages.
    
    Building upon these essential techniques for RBSs, a variety of optimal control problems can be formulated and efficiently solved by selecting appropriate optimization algorithms to enhance system performance from multiple perspectives. This will be further investigated in our future study. 
    
\vspace{-5pt}
    \bibliographystyle{IEEEtran}
    \bibliography{library_abbreviated}
    \newpage

    \vfill

\end{document}